\documentclass[prb, letterpaper, twocolumn, fleqn, floatfix, showpacs, showkeys ]{revtex4}

\usepackage{times, graphicx, units,mathrsfs,euscript,upgreek,amssymb,pifont}
\usepackage[normalem]{ulem}

\newcommand{\ve}{\epsilon}

\newcommand{\bv}{{\bf v}}

\newcommand{\br}{{\bf r}}
\newcommand{\bq}{{\bf q}}
\newcommand{\Phiind}{\Phi_\mathrm{ind}}
\newcommand{\FPhiind}{\FPhi_{\mathrm{ind}}}
\newcommand{\FPhi}{\widetilde{\Phi}}
\newcommand{\Fsig}{\widetilde{\sigma}}
\newcommand{\Phiext}{\Phi_\mathrm{ext}}
\newcommand{\FPhiext}{\FPhi_{\mathrm{ext}}}

\begin{document}

\title{High-energy plasmon spectroscopy of freestanding multilayer graphene}

\author{V. Borka Jovanovi\'{c}}
\author{I. Radovi\'{c}}
\author{D. Borka}
\affiliation{Vin\v{c}a Institute of Nuclear Sciences, University of
Belgrade, P.O. Box 522, 11001 Belgrade, Serbia}
\author{Z. L. Mi\v{s}kovi\'{c}}
\email{zmiskovi@math.uwaterloo.ca}
\affiliation{Department of Applied Mathematics, University of Waterloo, Waterloo, Ontario, Canada N2L 3G1}

%\date{\today}
\date{13 October 2011}



\begin{abstract}
We present several applications of the layered electron gas model to
electron energy loss spectroscopy of free-standing films consisting
of $N$ graphene layers in a scanning transmission electron
microscope. Using a two-fluid model for the single-layer
polarizability, we discuss the evolution of high-energy plasmon
spectra with $N$, and find good agreement with the recent
experimental data for both multi-layer graphene with $N<10$, and
thick slabs of graphite. Such applications of these analytical
models help shed light on several features observed in the plasmon
spectra of those structures, including the role of plasmon
dispersion, dynamic interference in excitations of various plasmon
eigenmodes, as well as the relevance of the bulk plasmon bands,
rather than surface plasmons, in classifying the plasmon peaks.
\end{abstract}

\maketitle
\thispagestyle{plain}

\section{Introduction}

Studying the relation between the plasmon spectra in thick graphite
samples and those in carbon nanostructures with one or several
layers has recently come to focus in carbon research. In that
context, multilayer graphene (MLG) offers a particularly suitable
system because of the precision that can be achieved in determining
the number of carbon layers $N$, as shown in the recent experimental
study of plasmon excitations in free-standing MLG by electron energy
loss spectroscopy (EELS) in scanning transmission electron
microscope (STEM). \cite{Gass_2008,Eberlein_2008} Those authors
found that the so-called low-loss EEL spectra are dominated by two
broad peaks that may be related to the $\pi$ and $\sigma+\pi$
plasmons of graphite, with the peak positions that change as the
number $N$ of layers in the MLG increases.
\cite{Gass_2008,Eberlein_2008} Similarly, plasmon spectra of
graphene were also discussed in the context of recent experimental
studies of thick layers of highly oriented pyrolytic graphite (HOPG)
by ultrafast electron microscopy (UEM) in TEM
\cite{Carbone_2009a,Carbone_2009b} and by inelastic x-ray scattering
(IXS), \cite{Hambach_2008,Reed_2010} as well as in the experiments
on multi-walled carbon nanotubes (MWCNTs) using both EELS and IXS.
\cite{Upton_2009} Given that in some of those studies the observed
dependence of plasmon frequencies on the number of layers has been
tentatively described in terms of the familiar concept of surface
and bulk plasmons, it is desirable to adopt an analytically
tractable theoretical model that can tackle this issue in an
explicit and transparent manner.

We demonstrate that a dielectric-response approach based on the
layered electron gas (LEG) model, where each layer of carbon atoms
is viewed as a two-dimensional electron gas (2DEG) of zero
thickness, \cite{Visscher_1971,Fetter_1974} provides useful insight
in the effect of the number of layers on the plasmon spectra of MLG.
The applicability of the LEG model to studying the high-frequency
plasmon modes in MLG by means of EELS is justified for typical
experimental conditions in STEM, where the target excitation is
dominated by the in-plane transfer of momentum $\bq$ of a fast
incident electron that traverses a sequence of graphene layers,
\cite{Gass_2008,Eberlein_2008,Carbone_2009a,Carbone_2009b} while
electron hopping between those layers may be neglected for
excitation energies that exceed the interlayer coupling of $t_\perp
\approx 0.4$ eV. \cite{Reed_2010} The LEG model has a long history
in the study of plasmon excitations in various structures, including
semiconductor superlattices,
\cite{Giuliani_1983,Jain_1985,Gumbs_1988} graphite intercalated
compounds (GICs), \cite{Shung_1986} high-$T_c$ superconductors,
\cite{Kresin_1988,Morawitz_1993} and MWCNTs.
\cite{Yannouleas_1996,Chung_2007} A similar model has also been
recently discussed in Ref.~\cite{Reed_2010} as a promising
theoretical tool for extracting spectroscopic information on
single-layer graphene (SLG) from their IXS data for thick graphite
samples. Given the increasing number of recent developments in
high-energy plasmon spectroscopy of graphene-based structures,
\cite{Gass_2008,Eberlein_2008,Carbone_2009a,Carbone_2009b,Hambach_2008,Reed_2010,Upton_2009}
one is led to a conclusion that revisiting the old LEG model is well
warranted. Due to its physical transparency and analytical
flexibility, the LEG model offers new insight into various
experimental observations that complements the insight based on the
results of computationally intensive \textit{ab initio} approaches.

The authors of Ref.~\cite{Eberlein_2008} provided a theoretical
discussion of their data by means of \textit{ab initio} calculations
of the loss function for MLG in the optical limit, $\bq=\mathbf{0}$,
based on the work of Marinopoulos \textit{et
al.},\cite{Marinopoulos_2004} where HOPG was modeled by using a
supercell of carbon layers separated by an equilibrium distance of
$d\approx$ 3.35 \AA. In Ref.~\cite{Eberlein_2008} an isolated SLG
and isolated MLGs with $N$ = 2 and 3 layers, having an interlayer
separation $d$, were simulated by using supercells where those
structures were periodically repeated with a separation between them
taken to be a multiple of $d$ (typically fivefold). As a consequence
of using supercells, the treatment of interlayer Coulomb interaction
in \textit{ab initio} calculations for the thus modeled SLG and MLGs
is necessarily approximate, giving rise to plasmon peaks in the loss
function whose intensities and the ``precise peak positions depend
on the separation'' that was adopted within a supercell, as noted by
the authors of Ref.~\cite{Eberlein_2008}. Nevertheless, the
\textit{ab initio} calculations gave good qualitative agreement with
the experimental EEL spectra for a SLG and MLGs with $N$ = 2 and 3
layers. \cite{Eberlein_2008} Moreover, that agreement was taken as
an indication that the optical limit of a loss function suffices for
modeling the EEL spectra, even though it was noted that those
spectra were recorded so that the wave vector ``$\bq$ has a
considerable in-plane component.'' \cite{Eberlein_2008} Namely,
while the 100 keV electrons traverse the MLG targets undergoing
negligible momentum transfer, the relatively large collection
semiangle of 19 mrad in STEM ensured that the EEL spectra were
recorded as being integrated over wave numbers up to $q_c\approx$
3.2 \AA$^{-1}$. \cite{Eberlein_2008} Furthermore, the authors of
Ref.~\cite{Eberlein_2008} did not consider any effects of the
incident electron trajectory, even though such effects may give rise
to a significant interference in plasmon excitations at different
carbon layers, which is critically dependent on proper use of the
interlayer separation. Finally, we mention that elements of
phenomenological modeling of the EEL spectra are introduced in the
\textit{ab initio} calculations by their use of a spectral
broadening of 1.5 eV to smooth out the resulting theoretical
spectra, and hence improve comparison with the experiment.
\cite{Eberlein_2008}

A possibly interesting alternative theoretical discussion of the
experimental EEL spectra \cite{Eberlein_2008} could be based on a
continuum dielectric model, which was used with notable success for
multilayer carbon nanostructures, different from the MLG, by
treating them as slabs of finite thickness described by a
frequency-dependent dielectric tensor in the optical limit.
\cite{Lucas_1995,Kociak_2000,Taverna_2002,Stephan_2002} However, as
pointed in Ref.~\cite{Eberlein_2008}, most applications of such an
anisotropic dielectric slab (ADS) model were restricted to the EELS
of curved nanostructures with nonpenetrating, or aloof electron
trajectories in STEM, so that excitations of the bulk plasmon modes
were suppressed with respect to the surface modes.
\cite{Lucas_1995,Kociak_2000,Taverna_2002,Stephan_2002} It should be
mentioned that the ADS model was used by Crawford
\cite{Crawford_1990} to study the stopping and deflection of swift
ions passing through the bulk of HOPG, but it is true that full
application of the ADS model to plasmon spectroscopy of MLG with
penetrating electron trajectories is yet to be undertaken. On the
other hand, while the ADS model is perceived as suitable for carbon
nanostructures with a large number of layers, difficulties arise if
one attempts to reach the limit of a one-atom-thick layer by
starting from a dielectric slab of finite thickness.
\cite{Stephan_2002} Namely, it was found that the slab thickness
strongly affects the coupling of surface plasmons at the opposite
sides of the slab, \cite{Perez_2010} and hence the authors remarked
that achieving the limit of a single-layer is not straightforward in
the ADS model because an ``arbitrary choice of the effective
dielectric thickness'' has to be made. \cite{Stephan_2002}

It is expected that the LEG model can provide useful additional
insight into the issues discussed in the above two paragraphs
regarding (a) the interlayer Coulomb interactions, (b) the effects
of large in-plane momentum transfer, (c) the dynamic interference
effect due to electron trajectory, (d) the lack of available
applications of the ADS model
\cite{Crawford_1990,Lucas_1995,Kociak_2000,Taverna_2002,Stephan_2002}
for penetrating electron trajectories, and (e) the single-layer
limit. Another benefit coming from the LEG model is revealed in the
limit of an infinite periodic lattice of identical layers, giving
rise to a particularly transparent description of the bulk plasmon
bands that may be of interest for studying the plasmon spectra of
HOPG. \cite{Giuliani_1983,Jain_1985,Shung_1986} Finally, it is
particularly convenient that the LEG model involves separate
treatments of the \textit{interlayer} Coulomb interactions and the
\textit{intralayer} dynamics by using a wave vector-dependent,
non-interacting polarizability function $\chi_0(\bq,\omega)$ of SLG
as an independent input quantity. Obviously, whether fine details in
the EEL spectra can be successfully modeled depends on the
availability of a good-quality single-layer polarizability
$\chi_0(\bq,\omega)$, but the effects of the increasing number of
layers within MLG are expected to be robustly reproduced by the LEG
model.

Over the past few years, sophisticated models have been developed
for $\chi_0(\bq,\omega)$ that are appropriate for describing the
\textit{low-energy} excitations (up to, say, 1 - 2 eV)
\cite{Nagashima_1992} in the vicinity of the \emph{K} points in the
Brillouin zone of SLG, where the conduction and valence $\pi$
electron bands may be approximated by Dirac cones with zero gap.
\cite{Castro_2009,Hwang_2010,Sarma_2011} The intra-band $\pi$
plasmon excitations (sometimes called sheet plasmons) were probed in
this energy range by means of high-resolution reflection EELS
(HREELS) of epitaxial graphene under heavy doping conditions,
\cite{Nagashima_1992,Liu_2008,Langer_2010,Liu_2010,Langer_2011}
giving dispersion relations that were studied theoretically by means
of $\chi_0(\bq,\omega)$, which included the effects of plasmon
damping \cite{Allison_2009} and plasmon-phonon coupling.
\cite{Allison_2010,Hwang_2010} However, plasmon excitations at such
low energies are hardly accessible in the TEM experiments because of
the presence of the zero loss peak that masks the spectral features
up to about 2 eV. \cite{Egerton_2009} On the other hand, high-energy
plasmon spectra, which were observed in thick graphite samples by
means of EELS, may be associated with the inter-band excitations of
both the $\pi$ and $\sigma$ electron bands with the gaps of about 4
eV and 14 eV, respectively (see Ref.~\cite{Marinopoulos_2004} and
the references therein). As far as the modeling of such high-energy
spectra in SLG is concerned, we note that the few recent
improvements of \textit{ab initio} calculations of the optical
dielectric function of graphene did not quite agree with regard to
the role of the excitonic effects associated with the $\sigma$
electron bands. \cite{Trevisanutto_2010,Chen_2011,Yang_2011}
Therefore, there is some value in the transparency offered by a
phenomenological two-fluid model for high-energy excitations of the
$\pi$ and $\sigma$ electrons in carbon. \cite{Cazaux_1970}
Accordingly, we adopt here such a model for SLG, while being aware
that subtle features due to low-energy, intra-band $\pi$ electron
excitations are inaccessible in the EELS anyway. \cite{Egerton_2009}
A tradeoff to using a phenomenological $\chi_0(\bq,\omega)$ as an
input to the LEG model is that the resulting analytical tractability
may help reveal how the main features in the EEL spectra of MLG,
i.e., the $\pi$ and $\sigma+\pi$ plasmon peaks, evolve as the number
of layers increases from $N=1$ in SLG to $N\rightarrow\infty$ in
HOPG. In particular, the LEG model will enable us to analyze the
role played in the spectra by the formation of the bulk plasmon
bands in HOPG, and to show that the concept of surface plasmons has
limited applicability in the present context.

After presenting theoretical details of the LEG model in the
following section, we discuss the comparison of our calculations
with several experiments using a phenomenological model for the
single-layer polarizability, which is followed by our concluding
remarks. Note that we use Gaussian electrostatic units, and we set
$\hbar=1$.

\section{The layered electron gas model}

In a typical STEM experiment operating at the voltage of 100 kV,
\cite{Eberlein_2008} the momentum transfer of the incident electron
is close to zero, so we shall use a straight-line trajectory while
neglecting relativistic effects. \cite{Egerton_2009} We use a
Cartesian coordinate system $\{\br,z\}$ with $\br=\{x,y\}$ and
assume that graphene layers occupy planes $z_n=(n\!-\!1\!) d$, where
$n=1,2,\ldots,N$ and $d$ is the interlayer spacing. The induced
potential in the system, $\Phiind(\br,z,t)$, may be expressed via
its Fourier transform with respect to the in-plane coordinates
($\br\rightarrow\bq$) and time ($t\rightarrow\omega$) as
\begin{eqnarray}
\FPhiind(\bq,z,\omega)= \sum_{n=1}^N\frac{2\pi}{q}\,\Fsig_n(\bq,\omega)\mathrm{e}^{-q|z-z_n|},
\label{Phiind}
\end{eqnarray}
where  $\Fsig_n(\bq,\omega)$ is the Fourier transform of the induced
charge density (per unit area) on the $n$th layer, which may be
written in a self-consistent field approximation as
\begin{eqnarray}
\Fsig_n(\bq,\omega)=-e^2\chi_0(\bq,\omega)\left[\FPhiext(\bq,z_n,\omega)+\FPhiind(\bq,z_n,\omega)\right],
\label{SCFA}
\end{eqnarray}
with $\Phiext(\br,z,t)$ being the external potential. From the
charge density of the incident electron,
$\rho_\mathrm{ext}(\br,z,t)=Ze\,\delta\!\left(\br\!-\!\bv_{\parallel}t\right)\delta\!\left(z\!-\!v_{\perp}t\right)$,
where $Z=-1$ and $\bv_{\parallel}$ and $v_{\perp}$ are the velocity
components parallel and perpendicular to the graphene planes,
respectively, we find
\begin{eqnarray}
\FPhiext(\bq,z,\omega)=\frac{4\pi Ze v_{\perp}}{\left(qv_{\perp}\right)^2+\left(\omega\!-\!\bq\cdot\bv_{\parallel}\right)^2}
%K(q,\omega-\bq\cdot\bv_{\parallel})
\,\mathrm{e}^{i\left(\omega-\bq\cdot\bv_{\parallel}\right) z/v_{\perp}}.
\label{Phiext}
\end{eqnarray}
Thus, by using Eqs.~(\ref{Phiind}) and (\ref{Phiext}) in
Eq.~(\ref{SCFA}), we obtain a matrix equation for
$\Fsig_n(\bq,\omega)$,
\begin{equation}
\begin{array}{l}
\sum_{n'=1}^N\mathcal{M}_{nn'}(\bq,\omega)\Fsig_{n'}(\bq,\omega) \\
= -ZeV(q)\chi(\bq,\omega)K(q,\omega\!-\!\bq\cdot\bv_{\parallel})
\psi_n(\omega\!-\!\bq\cdot\bv_{\parallel}),
\end{array}
\label{matrix_eq}
\end{equation}
where
\begin{eqnarray}
\mathcal{M}_{nn'}(\bq,\omega)=\delta_{nn'}+\left(1-\delta_{nn'}\right)V(q)\chi(\bq,\omega)\mathrm{e}^{-qd|n'-n|},
\label{matrix}
\end{eqnarray}
\begin{eqnarray}
K(q,\omega)=\frac{2q v_{\perp}}{\left(qv_{\perp}\right)^2+\omega^2},
\label{K}
\end{eqnarray}
\begin{eqnarray}
\psi_n(\omega)=\mathrm{e}^{i(n-1)\omega d/v_{\perp}},
\label{psi}
\end{eqnarray}
and $\chi(\bq,\omega)=\chi_0(\bq,\omega)/\ve(\bq,\omega)$, where
$\ve(\bq,\omega)=1+V(q)\chi_0(\bq,\omega)$ is the intra-layer
dielectric function of SLG with $V(q)=2\pi e^2/q$ being the Fourier
transformed Coulomb interaction in two dimensions (2D).

In the next step we solve Eq.~(\ref{matrix_eq}) for
$\Fsig_n(\bq,\omega)$ by inverting the matrix $\mathcal{M}$ with
elements given in Eq.~(\ref{matrix}). Substituting this solution
into Eq.~(\ref{Phiind}) enables one to express $\Phiind(\br,z,t)$
via an inverse Fourier transform, so that the total energy lost by
the incident electron may be evaluated from \cite{Mowbray_2010}
\begin{eqnarray}
E_\mathrm{loss} = -\int\limits_{-\infty}^\infty dt\int d^2\br\int\limits_{-\infty}^\infty dz\,\rho_\mathrm{ext}(\br,z,t)
\,\frac{\partial}{\partial t}\Phiind(\br,z,t).
 \label{Eloss}
\end{eqnarray}
Using symmetry properties of the function $\chi_0(\bq,\omega)$ at
zero temperature, one may write  Eq.~(\ref{Eloss}) as
\begin{eqnarray}
E_\mathrm{loss} \label{Eloss_P}=\int\limits_0^\infty d\omega\,\omega P_N(\omega),
\end{eqnarray}
where $P_N(\omega)$ is the probability density for loosing the
energy $\omega$, given by

\begin{equation}
\begin{array}{l}
P_N(\omega)= \\
\frac{(Ze)^2}{2\pi^2}\int \frac{d^2\bq}{q}\, K^2(q,\omega\!-\!\bq\cdot\bv_{\parallel})
\Im\left[V(q)\chi(\bq,\omega)Q_N(\bq,\omega)\right],
\end{array}
\label{P}
\end{equation}

\noindent with
\begin{equation}
\begin{array}{l}
Q_N(\bq,\omega)= \\
\sum_{n=1}^N \sum_{n'=1}^N \psi_n^*(\omega\!-\!\bq\cdot\bv_{\parallel}) \left(\mathcal{M}^{-1}\right)_{nn'}
\psi_{n'}(\omega\!-\!\bq\cdot\bv_{\parallel}).
\end{array}
\label{Q}
\end{equation}

\noindent Note that the density $P_N(\omega)$ will be directly
compared with the experimental EEL spectra of MLG with finite $N$,
which were taken under the normal electron incidence.
\cite{Eberlein_2008} So, setting $\bv_{\parallel}=\mathbf{0}$ in
Eqs.~(\ref{P}) and (\ref{Q}) and invoking the near-isotropy of
graphene's polarizability, \cite{Marinopoulos_2004}
$\chi_0(\bq,\omega)=\chi_0(q,\omega)$, renders the angular
integration in Eq.~(\ref{P}) trivial. The remaining integration over
the wave numbers should go up to $q_c\approx$ 3.2 \AA$^{-1}$,
\cite{Eberlein_2008} but we found that no difference occurs in the
final results for $P_N(\omega)$ if the upper limit is extended to
$\infty$ because the kinematic factor $K^2(q,\omega)$ in
Eq.~(\ref{P}) is strongly peaked at $q=\omega/v_\perp\ll q_c$ for
frequencies of interest here, cf.\ Eq.~(\ref{K}).

Further note that the factor $\left(\mathcal{M}^{-1}\right)_{nn'}$
in Eq.~(\ref{Q}) is a $(q,\omega)$ dependent element of a matrix
$\mathcal{M}^{-1}$ that is inverse to the matrix $\mathcal{M}$
defined in Eq.~(\ref{matrix}). Thus, the quadratic form
$Q_N(q,\omega)$ in Eq.~(\ref{Q}) can be relatively easily obtained
using symbolic computation software for, say, $N<10$. For example,
whereas for a single-layer $Q_1(q,\omega)=1$, we obtain from
Eq.~(\ref{Q}) for a bilayer graphene
\begin{eqnarray}
Q_2(q,\omega)=2\frac{1-V(q)\chi(q,\omega)\mathrm{e}^{-qd}\cos\left(\omega d/v_{\perp}\right)}{1-V^2(q)\chi^2(q,\omega)\mathrm{e}^{-2qd}}.
 \label{Q2}
\end{eqnarray}

While the analytical results for $Q_N$ become increasingly
cumbersome with increasing $N$, we note that a relatively simple
expression for $P_N$ may be obtained in the limit
$N\rightarrow\infty$, which we shall denote by $P_\infty$,
corresponding to an electron traversing a sufficiently thick slab of
HOPG, such that the end effects may be neglected if $Ndq\gg 1$. In
that case, the matrix equation in Eq.~(\ref{matrix_eq}) may be
solved by using Fourier series with a wavenumber $k$ in the
direction perpendicular to the graphene planes, giving

\begin{equation}
\begin{array}{r}
P_\infty(\omega)=N \frac{(Ze)^2}{2\pi^2}\int \frac{d^2\bq}{q}\, K^2(q,\omega-\bq\cdot\bv_{\parallel})\, \\
\times \Im\!\left[\frac{V(q)\chi_0(q,\omega)}{1+S\!\left(q,\frac{\omega-\bq\cdot\bv_{\parallel}}{v_\perp}\right) V(q)\chi_0(q,\omega)}\right],
\end{array}
\label{P_bulk}
\end{equation}

\noindent where

\begin{eqnarray}
S(q,k)=\frac{\sinh(qd)}{\cosh(qd)-\cos(k d)}
 \label{S}
\end{eqnarray}

\noindent is the Coulomb structure factor for an infinite periodic
lattice of layers. \cite{Fetter_1974} Note that the density
$P_\infty(\omega)$ will be directly compared with the experimental
EEL spectra of HOPG, which were taken under the normal electron
incidence, \cite{Carbone_2009a} so we again set
$\bv_{\parallel}=\mathbf{0}$ in Eq.~(\ref{P_bulk}).

Finally, in order to identify the exact nature of plasmons giving
the most prominent contributions to the spectral densities
$P_N(\omega)$ or $P_\infty(\omega)$, it is worthwhile analyzing the
eigenmodes of the underlying MLG, which are obtained by setting the
damping rates in $\chi_0(q,\omega)$ to zero. Thus, for plasmon modes
in the case of finite $N$, one has to solve the equation
$\ve^N(q,\omega)\det\left(\mathcal{M}\right)=0$, giving frequencies
at which the factor
$\Im\left[V(q)\chi(q,\omega)Q_N(\bq,\omega)\right]$ in Eq.~(\ref{P})
becomes singular. On the other hand, eigenmodes in an infinite
periodic lattice of 2DEG layers are obtained by solving the equation
\begin{eqnarray}
1+S\!\left(q,k\right) V(q)\chi_0(q,\omega)=0,
 \label{infinite}
\end{eqnarray}
with $k$ as a parameter. \cite{Giuliani_1983,Jain_1985}  By letting
$0\le k\le\pi/d$ and using a one-fluid model for $\chi_0(q,\omega)$,
one obtains a band of dispersion relations for the so-called bulk
plasmon modes that propagate with the wave numbers $q$ parallel to
the layers. \cite{Fetter_1974,Giuliani_1983,Jain_1985,Gumbs_1988}
Similarly, by using a two-fluid model for $\chi_0(q,\omega)$ in
Eq.~(\ref{infinite}), one obtains two such bands for $0\le
k\le\pi/d$ that correspond to the $\pi$ and $\sigma +\pi$ plasmons
in the bulk of HOPG. In the case of a semi-infinite lattice of
equally spaced identical layers of 2DEG, it was shown that a plasmon
mode may arise with the dispersion relation outside the plasmon
band(s), only if there is a mismatch between the background
dielectric constants in the lattice and the nearby space.
\cite{Giuliani_1983, Jain_1985} Hence, this kind of surface plasmon,
which is localized near the boundary layer of a semi-infinite
lattice, is not expected to exist in the $N\rightarrow\infty$ limit
of the LEG model for HOPG placed in vacuum or air.

\section{Results and discussion}

A three-dimensional (3D) version of the phenomenological two-fluid
polarizability function was used in the ADS model to build a
dielectric tensor in the optical limit with suitable Drude-Lorentz
parameters \cite{Lucas_1995} for modeling of the EEL spectra of
multilayer fullerene molecules \cite{Kociak_2000} and MWCNTs.
\cite{Taverna_2002,Stephan_2002} In other applications, such a 3D
version of the two-fluid model was used to study the variable degree
of the $sp^2$ hybridization for applications in different carbon
materials \cite{Calliari_2007} and the in-plane plasmons in HOPG,
\cite{Calliari_2008} as well as to deduce the optical conductivity
of graphene in order to calculate Casimir forces between graphene
layers. \cite{Drosdoff_2010} However, we need here a strictly 2D
version of the two-fluid model with suitable Drude-Lorentz
parameters, similar to that used to describe plasmon excitations in
single-layer fullerene molecules \cite{Barton_1993,Gorokhov_1996}
and single-wall carbon nanotubes (SWCNTs). \cite{Jiang_1996} One way
of \textit{deriving} such a polarization function for SLG could
proceed from a 2D, two-fluid hydrodynamic model with Thomas-Fermi
and Dirac (TFD) interactions, \cite{Mowbray_2004,Mowbray_2010} which
enabled a semiquantitative comparison with the plasmon dispersion
relations that were observed in the EEL spectra of SWCNTs over a
broad range of wavelengths in the axial direction.
\cite{Kramberger_2008}

We note that clear distinction should be made between the 3D and 2D
Drude-Lorentz models in the sense that the former class of models
usually employs only frequency-dependent dielectric functions,
whereas the latter class necessarily involves nonlocal effects,
i.e., a dependence on the in-plane wave number that reflects
incomplete Coulomb screening by an electron gas in 2D.
\cite{Perez_2010} A formal connection between the two types of
models may be established by considering a thin film, where both
symmetric and antisymmetric coupling of surface plasmons at the
opposite sides of the film arise, in addition to the bulk plasmon
modes. Keeping in mind that, in realistic applications to one-atom
thick layers, only the surface density, $n$, of electrons may be
defined unambiguously, taking the zero-thickness limit of a film
leaves the lower-energy, symmetric, in-plane surface plasmon as the
only observable excitation mode, characterized by a typical
square-root plasmon dispersion of the form $\omega \sim\sqrt{nq}$ in
a quasifree 2DEG. \cite{Perez_2010} Whether such a plasmon mode
should be referred to as a surface plasmon, or simply an intrinsic
plasmon mode of a 2DEG is a matter of semantics when it comes to
one-atom-thick layers.

Hence we adopt from Ref.~\cite{Mowbray_2010} a planar version of the
2D, two-fluid hydrodynamic model that gives
$\chi_0=\chi_\sigma^{(0)}+\chi_\pi^{(0)}$ for SLG, where
\begin{eqnarray}
\chi_\nu^{(0)}(q,\omega)=\frac{n_\nu^0q^2/m_\nu^*}{s_\nu^2 q^2 +\omega_{\nu r}^2-\omega\left(\omega+i\gamma_\nu\right)},
\label{chi}
\end{eqnarray}
with $n_\nu^0$, $m_\nu^*$, $s_\nu$, $\omega_{\nu r}$, and
$\gamma_\nu$ being the equilibrium surface number density of
electrons, effective electron mass, acoustic speed, restoring
frequency, and the damping rate in the $\nu$th fluid (where
$\nu=\sigma, \pi$), respectively. Note that the restoring
frequencies for the in-plane electron excitations are related to the
$\pi\rightarrow\pi^*$ and $\sigma\rightarrow\sigma^*$ inter-band
transitions, \cite{Barton_1993} which were found to dominate the
in-plane loss function of SLG in the optical limit at energies close
to 4 and 14 eV, respectively. \cite{Marinopoulos_2004} On the other
hand, the terms involving the acoustic speeds in Eq.~(\ref{chi})
arise from the TFD interactions in the hydrodynamic model,
\cite{Mowbray_2010} but their contribution to the EEL spectra turns
out to be negligible in the present context because the kinematic
factor $K^2(q,\omega)$ in Eqs.~(\ref{P}) and (\ref{P_bulk}) is
strongly peaked at $q=\omega/v_\perp$ and because $s_\nu\ll
v_\perp$.

\begin{figure}
\centering
\includegraphics[width=0.48\textwidth]{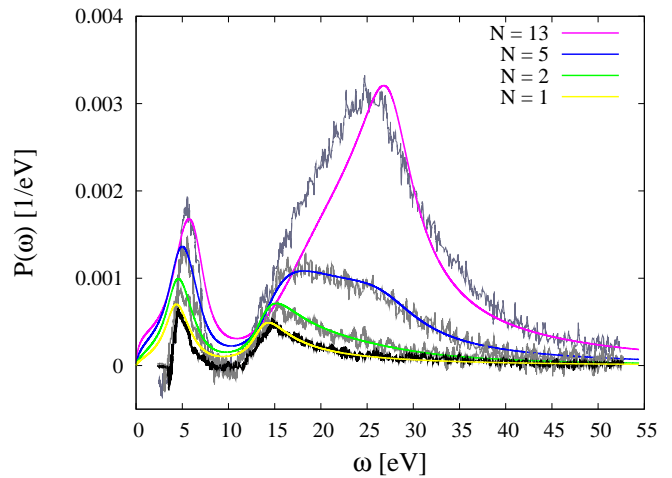}
\caption{(Color online.) Probability density $P_N(\omega)$ (in 1/eV) versus energy loss $\omega$ (in eV), evaluated from Eq.\ (\ref{P}) for $N$
= 1, 2, 5, and 13 graphene layers [smooth solid (yellow, green, blue, and pink) curves], along with the corresponding experimental EEL spectra
from Ref.~\cite{Eberlein_2008} [noisy (gray) curves].}
\label{Fig1}
\end{figure}

Taking $m_\nu^*$ to be the free electron mass and using the
unperturbed surface electron densities of SLG, $n_\pi^0\approx$ 38
nm$^{-2}$ and $n_\sigma^0=3n_\pi^0\approx$ 115 nm$^{-2}$, we treat
the remaining parameters in Eq.~(\ref{chi}) as adjustable. The best
fit to the experimental EEL spectra \cite{Eberlein_2008} is found
for $\omega_{\pi r}=4.08$ eV, $\omega_{\sigma r}=13.06$ eV,
$\gamma_\pi=2.45$ eV, and $\gamma_\sigma=2.72$ eV. The results for
$P_N(\omega)$ were thus computed from Eq.\ (\ref{P}) with $N$ = 1,
2, 5 and 13, and are compared in Fig.\ref{Fig1} with the
experimental curves from Fig.\ 1(e) of Ref.\ \cite{Eberlein_2008},
corresponding to 1, 2, 5, and $>\!\!10$ graphene layers (We found
that our curve $N=13$ provides the best fit with their curve labeled
 ``$>\!$10L''.). We note that the experimental curves were taken
under the same acquisition conditions, thereby enabling a direct
quantitative comparison among them. \cite{Eberlein_2008} On the
other hand, apart from adjusting the arbitrary unit of the
experimental curves to the absolute unit of our $P_N(\omega)$, no
relative scaling took place among the theoretical curves with
different numbers of layers. One notes that the experimental curves
are well reproduced by the LEG model for energies $\omega \gtrsim$ 3
eV and $N<10$, both in magnitude and in the shape of spectra.
Regarding the discrepancy at $\omega \lesssim$ 3 eV, we note that
the present version of the two-fluid model in Eq.~(\ref{chi}), along
with the neglect of interlayer tunneling, is only expected to work
in MLG for high-energy excitations, but also that the complete
vanishing of the experimental spectra at energies below 3 - 4 eV may
be a consequence of the method used in their subtraction of the
zero-loss peak. \cite{Eberlein_2008}

Furthermore, while the above values of the parameters $\omega_{\pi
r}$, $\omega_{\sigma r}$, $\gamma_\pi$, and $\gamma_\sigma$ are
quite close to those listed in the Table 2 of Ref.~\cite{Lucas_1995}
for the in-plane dielectric function of a 3D, two-fluid model of
ADS, features seen in the spectra in Fig.\ref{Fig1} are relatively
robustly reproduced by theoretical curves for other choices of these
parameters. In particular, the damping rates are strongly affected
by the presence of impurities or defects, which serve as scattering
centers for charge carriers in individual carbon layers. While for
graphene on a substrate the concentration of impurities strongly
varies from sample to sample, the freestanding graphene may be
relatively clean in the case of one or few layers, but thicker
samples may contain increased amounts of impurities and defects.
Thus, assigning smaller values, or possibly allowing for frequency-
dependent damping rates $\gamma_\pi$ and $\gamma_\sigma$ as in Ref.\
\cite{Djurisic_1999}, could improve the agreement for energies
around 10 eV in Fig.~\ref{Fig1}, where the experiment exhibits an
almost complete depletion of the spectra. On the other hand, the
width of the high-energy peak at about 27 eV for $N>10$ is not well
reproduced by the present choice of parameters, but agreement may be
improved if one allows for higher damping rates due to increased
density of impurities. \cite{Eberlein_2008} This point will be taken
up later in the discussion of Fig.~\ref{Fig6}.

The most important trends seen in Fig.~\ref{Fig1} are that the $\pi$
plasmon peak position moves from about 5 eV to about 7 eV as $N$
increases without significant changes in its shape, whereas the
$\sigma+\pi$ plasmon peak at about 15 eV for $N=1$ evolves through
the development of a plateau between 15 and 27 eV for $N=5$, to be
dominated by a peak at about 27 eV for $N=13$, with a growth in
magnitude that exceeds the growth of the $\pi$ plasmon peak. The
lowest peak positions that occur for $N=1$ and the highest peak
positions that occur for $N>10$ in Fig.~\ref{Fig1} were associated
in Ref.~\cite{Eberlein_2008} with the surface and bulk plasmon
frequencies of graphite, respectively, for both the $\pi$ and
$\sigma+\pi$ plasmons. Similarly, we shall see in Fig.~\ref{Fig6}
that the $\sigma+\pi$ plasmon contributions at about 15 and 27 eV in
the EEL spectra of HOPG were also associated in
Ref.~\cite{Carbone_2009a} with the surface and bulk plasmons,
respectively. In the following, we shall use the LEG model to
discuss how adequate those associations are.

\begin{figure}
\centering
\includegraphics[width=0.48\textwidth]{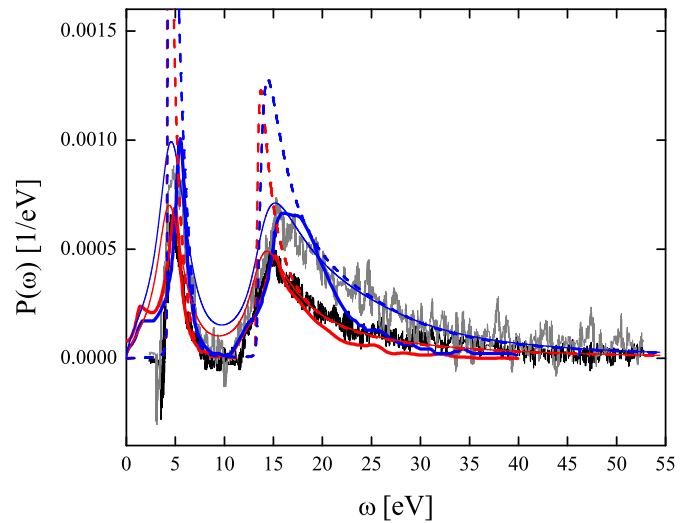}
\caption{(Color online.) Probability density $P_N(\omega)$ (in 1/eV) versus energy loss $\omega$ (in eV), evaluated from Eq.\ (\ref{P}) for $N$
= 1 [lower (red) curves] and 2 [upper (blue) curves] graphene layers with non-zero damping (smooth thin solid lines) and vanishing damping
(dashed lines), along with the corresponding experimental data [noisy (grey) curves] and the \textit{ab initio} curves (smooth thick solid
lines) from Ref.~\cite{Eberlein_2008}.}
\label{Fig2}
\end{figure}

In Fig.~\ref{Fig2} we analyze the details of the experimental EEL
spectra of Eberlein \textit{et al.}~for $N=1$ and 2, corresponding
to a SLG and a bilayer graphene (BLG), which are reproduced from
Fig.~\ref{Fig1} (noisy curves) along with the theoretical curves
from the LEG model for $P_1(\omega)$ and $P_2(\omega)$ from
Eq.~(\ref{P}) (thin solid lines) and with two theoretical curves,
taken from Fig.~3(b) of Ref.~\cite{Eberlein_2008} (thick sold
lines), which are based on their \textit{ab initio} calculations.
The most prominent feature of the experimental curves is the
asymmetry in the $\sigma+\pi$ plasmon peaks, which is well
reproduced by the LEG model, but not by the \textit{ab initio}
calculations by Eberlein \textit{et al.}\cite{Eberlein_2008} (We
note, however, that such asymmetry was observed in the \textit{ab
initio} calculations based on a \textit{GW} and Bethe-Salpeter
equation approach by Trevisanutto \textit{et al.}
\cite{Trevisanutto_2010}) We emphasize that the long tails in those
peaks, which extend at energies beyond 15 eV, are a consequence of
the integration over $q$ in Eq.~(\ref{P}) that captures the effect
of dispersion of the $\sigma+\pi$ plasmon, which is neglected when
optical limit of the loss function is used. To further elucidate
this point, we have recalculated the curves for $P_1(\omega)$ and
$P_2(\omega)$ from Eq.~(\ref{P}) using the same parameters of the
two-fluid model in Eq.~(\ref{chi}), except that the damping rates of
both $\sigma$ and $\pi$ electrons were made very small. The
resulting curves are displayed in Fig.~\ref{Fig2} by the dashed
lines, showing a steep raise at the restoring frequencies
$\omega_{\nu r}$, along with the tails that extend for
$\omega>\omega_{\nu r}$, giving a very good approximation to the
experimental curves at those frequencies.

\begin{figure}
\centering
\includegraphics[width=0.48\textwidth]{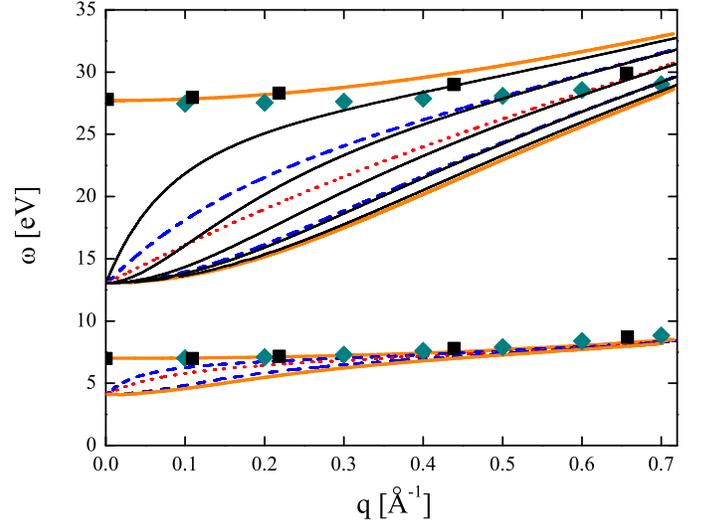}
\caption{(Color online.) Dispersion curves for the $\pi$ (lower group) and $\sigma+\pi$ (upper group) plasmons in MLG, obtained from the LEG
model with $N=1$ [dotted (red) lines] and $N=2$ [dashed (blue) lines], as well as $N=5$ [only the upper group is shown by (black) solid lines].
The upper and the lower edges are shown for both the $\pi$ and $\sigma+\pi$ plasmon bands in HOPG [thick (orange) solid lines]. The (dark cyan)
diamonds show the experimental plasmon peak positions in the EEL spectra of HOPG from Ref.~\cite{Kramberger_2010}, whereas the (black) squares
show the peak positions obtained by \textit{ab initio} calculations for graphite in Ref.~\cite{Marinopoulos_2004}.}
\label{Fig3}
\end{figure}

This observation may be easily rationalized for $N=1$ by referring
to the two-fluid model for $\chi_0(q,\omega)$ in Eq.\ (\ref{chi}) in
the limit $\gamma_\sigma=\gamma_\pi\rightarrow 0$, which was
discussed in the Appendix C to Ref.~\cite{Mowbray_2010}. When used
in Eq.~(\ref{P}) for $P_1(\omega)$, this limit converts the factor
$\Im\left[V(q)\chi(q,\omega)\right]$ into a superposition of two
$\delta$ functions, $\delta(\omega^2-\omega_\pm^2(q))$, where
$\omega_\pm(q)$ are the eigenfrequencies describing hybridization
between the $\sigma$ and $\pi$ electron fluids in a SLG, which are
obtained by solving the equation $\ve(q,\omega)=0$ as

\begin{equation}
\begin{array}{l}
  \label{pm}
  \omega^2_{\pm} = \frac{\omega_{\sigma}^2+\Omega_{\sigma}^2+\omega_{\pi}^2+\Omega_{\pi}^2}{2} \pm \sqrt{\left(\frac{\omega_{\sigma}^2+\Omega_{\sigma}^2-\omega_{\pi}^2-\Omega_{\pi}^2}{2}\right)^2+\Omega_{\sigma}^2\Omega_{\pi}^2},
\end{array}
\end{equation}

\noindent where we define

\begin{equation}
\omega_{\nu}(q)\equiv\sqrt{\omega_{\nu r}^2+s_\nu^2q^2}
\label{omega}
\end{equation}

\noindent and
\begin{equation}
\Omega_{\nu}(q)\equiv\sqrt{2\pi e^2n_\nu^0q/m_\nu^*}
\label{Omega}
\end{equation}

\noindent for the $\nu$th fluid. \cite{Mowbray_2010} The
eigenfrequencies $\omega_+(q)$ and $\omega_-(q)$ are labeled as the
$\sigma+\pi$ and the $\pi$ plasmon modes of SLG, and the
corresponding dispersion relations are shown in Fig.~\ref{Fig3} by
the upper and lower dotted lines, respectively. For the purpose of
the present discussion, it suffices to expand the expressions in
Eq.\ (\ref{pm}) to the first order in $q$, so that
$\omega_+^2\approx\omega_{\sigma r}^2+\Omega_{\sigma}^2(q)$ and
$\omega_-^2\approx\omega_{\pi r}^2+\Omega_{\pi}^2(q)$, describing
the long wavelength behavior of plasmon frequencies of the
non-interacting $\sigma$ and $\pi$ fluids in SLG. Using these
expressions for $\omega_\pm^2(q)$ in the delta functions
$\delta(\omega^2-\omega_\pm^2(q))$ makes the integration over $q$ in
Eq.~(\ref{P}) trivial, giving two contributions to $P_1(\omega)$ of
the form $\sim n_\nu^0/(\omega^2-\omega_{\nu r}^2)$, which
essentially capture the behavior of the tails seen in
Fig.~\ref{Fig2} for $\omega^2-\omega_{\nu r}^2\gg 2\pi e^2
n_\nu^0\omega/\left(m_\nu^*v_\perp\right)$.

Furthermore, it is interesting to comment on the similarity between
the experimental spectra for BLG and SLG, seen in Fig.~\ref{Fig2}.
In particular, their high-frequency tails at $\omega\gtrsim 20$ eV
seem to imply a scaling ratio of about 2:1, as remarked by Eberlein
\textit{et al.}. \cite{Eberlein_2008} This is best rationalized by
resorting to the eigenmode analysis for BLG by setting
$\gamma_\sigma=\gamma_\pi\rightarrow 0$ in Eq.~(\ref{P}) for
$P_2(\omega)$. Then, it can be shown that the factor
$\Im\left[V(q)\chi(q,\omega)Q_2(q,\omega)\right]$ becomes singular
at four frequencies, which are given by the expressions in
Eq.~(\ref{pm}) when frequencies $\Omega_{\nu}(q)$ in
Eq.~(\ref{Omega}) are replaced by
$\Omega_{\nu}^{\pm}(q)\equiv\Omega_{\nu}(q)\sqrt{1\pm\exp(-qd)}$.
The corresponding dispersion curves are shown in Fig.~\ref{Fig3} by
the dashed lines, two in the upper ($\sigma+\pi$ plasmon) group and
two in the lower ($\pi$ plasmon) group. However, the weakly
(quadratically) dispersing lower curves in each plasmon group
contribute to $P_2(\omega)$ with negligible spectral weights because
of the smallness of the factor $\omega d/v_{\perp}$ at the
frequencies of interest here. Namely, by setting $\cos\left(\omega
d/v_{\perp}\right)\approx 1$ in Eq.~(\ref{Q2}), one obtains
$Q_2(q,\omega)\approx 2/\left[1+V(q)\chi(q,\omega)\exp(-qd)\right]$,
so that the dominant spectral weight in $P_2(\omega)$ comes mostly
from the upper (linearly dispersing) curves in the $\sigma+\pi$ and
$\pi$ plasmon groups, seen in Fig.~\ref{Fig3} for BLG. This may be
further simplified by noting that the kinematic factor
$K^2(q,\omega)$ is strongly peaked at $q=\omega/v_\perp$, so that
$\exp(-qd)\approx 1$, and hence the factor
$\Im\left[V(q)\chi(q,\omega)Q_2(q,\omega)\right]$ in Eq.~(\ref{P})
for $P_2(\omega)$ becomes
$-\Im\left\{1/\left[1+2V(q)\chi_0(q,\omega)\right]\right\}$,
corresponding to a SLG with doubled $\sigma $ and $\pi$ electron
densities. Accordingly, the slopes of the upper dispersion curves
for BLG in the $\sigma+\pi$ and $\pi$ plasmon groups are seen in
Fig.~\ref{Fig3} to be about twice the slopes of the corresponding
dispersion curves for SLG in the limit of long wavelengths. As a
consequence, the peak positions in $P_2(\omega)$ occur at about the
same frequencies as those in $P_1(\omega)$, whereas the
high-frequency tails in $P_2(\omega)$ are approximated by $\sim
2n_\nu^0/(\omega^2-\omega_{\nu r}^2)$, confirming their ratio of 2:1
relative to the tails in $P_1(\omega)$.

The above analysis of SLG and BLG shows that various eigenmodes of
the underlying structure enter the EEL spectra with weights that are
strongly dependent on the incident electron speed. This dynamic
aspect of the LEG model, which is not covered in the \textit{ab
initio} calculations, \cite{Eberlein_2008} is related to the
interference effect in plasmon excitations at different carbon
layers within a MLG, and it may be quantified by the factor $\omega
T_N$ that appears in the expression for $Q_N(q,\omega)$, cf.\
Eqs.~(\ref{matrix}), (\ref{psi}) and (\ref{Q}), where $T_N\equiv
Nd/v_\perp$ is the time it takes the incident electron to traverse
the full thickness of MLG (neglecting relativistic effects).
Therefore, the interference effect will be negligible for fast
incident electrons in STEM at frequencies of interest in the
low-loss EELS of MLGs with a few layers giving $\omega T_N\ll 1$.
The smallness of the factor $\omega T_N=Nd\omega/v_\perp$ is
equivalent to the limit $d\rightarrow 0$, giving a picture where all
layers collapse onto a single sheet or, equivalently, the in-plane
plasmon modes in all layers oscillate in phase, which explains the
similarity among the spectra and the approximate scaling of their
high-energy tails as $P_N\propto N$ for $N\sim 1$. In other words,
the EEL spectra of MLGs with $N=2, 3,\ldots$ are dominated by the
peaks that are derived from the $\pi$ and $\sigma+\pi$ plasmons of
SLG, and hence they should not be associated with the surface
plasmons of graphite. \cite{Eberlein_2008}

\begin{figure}
\centering
\includegraphics[width=0.48\textwidth]{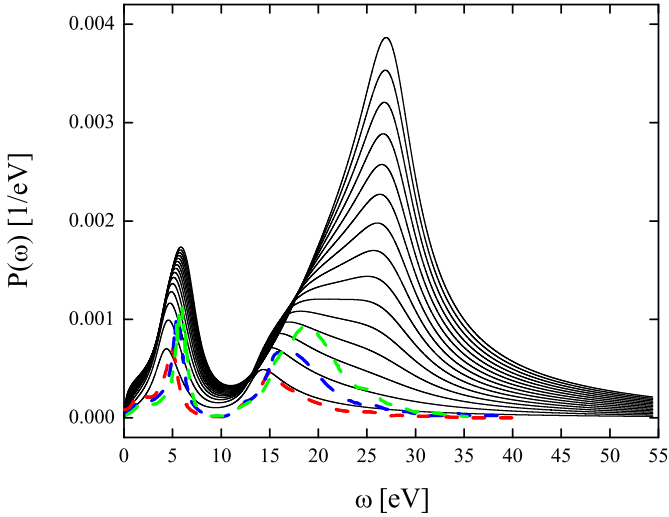}
\caption{(Color online.) Probability density $P_N(\omega)$ (in 1/eV) versus energy loss $\omega$ (in eV), evaluated from Eq.~(\ref{P}) for $N$
in the range from 1 to 15 with unit steps, using the same parameters as in Fig.~\ref{Fig1} [solid (black) curves]. Also shown are the results of
\textit{ab initio} calculations from Ref.~\cite{Eberlein_2008} for graphene with $N=1,2$, and 3 layers [dashed (red, blue, and green,
respectively) curves].}
\label{Fig4}
\end{figure}

When $N\gtrsim 10$, the factor $\omega T_N=Nd\omega/v_\perp$ may no
longer be small, and hence the interference effect may become
significant in sufficiently thick MLGs. Obviously, the onset of the
interference effect will occur at a smaller $N$ if the incident
electron moves at a lower speed. Moreover, because of the difference
in the relevant frequency ranges, one expects that the interference
effect will be more prominent for the $\sigma+\pi$ than for the
$\pi$ plasmons. This may explain why the $\sigma+\pi$ plasmon peak
in Fig.~\ref{Fig1} undergoes a more substantial change with
increasing $N$ than the $\pi$ plasmon peak. To illustrate this
point, we show in Fig.~\ref{Fig4} a detailed evolution of the EEL
spectra $P_N(\omega)$ from Eq.\ (\ref{P}) for $N$ going from 1 to 15
in unit steps, which were calculated with the same parameters of the
two-fluid model in Eq.~(\ref{chi}) as those used in Fig.~\ref{Fig1}.
(For the sake of further comparison with \textit{ab initio}
calculations, we also show in our Fig.~\ref{Fig4} the results from
Fig.~3(b) of Ref.~\cite{Eberlein_2008} for the in-plane spectra of
graphene with $N=1,2$, and 3 layers.) One confirms in
Fig.~\ref{Fig4} that in the LEG model the $\pi$ plasmon peak moves
continuously from about 5 to about 7 eV as $N$ increases, whereas
the spectral structure associated with the $\sigma+\pi$ plasmon
appears to evolve as a superposition of two peaks, one at about 15
eV and the other at about 27 eV, which do not move much as $N$
increases but rather have their weights strongly dependent on $N$.
One may further see from Fig.~\ref{Fig4} that the contribution from
the peak at about 15 eV dominates the spectra for $N\lesssim 5$, and
is still visible for $N\lesssim 10$, whereas the spectra with $N>10$
are dominated by the peak at about 27 eV. In particular, it appears
in Fig.~\ref{Fig4} that the weights of those two peaks are
approximately equal for MLG with $N=6$ layers, giving a value of the
relevant interference factor of $\omega T_6\approx 0.5$. Such a
peculiar composition of the $\sigma+\pi$ plasmon peak for $N\sim$ 5
- 6 deserves further analysis.

One expects that the plasmon eigenfrequencies in MLG with finite $N$
will fall into two disjoint groups, each giving $N$ dispersion
curves that correspond to the $\sigma+\pi$ and $\pi$ plasmons of
SLG. All the curves within each group become degenerate as
$q\rightarrow 0$ and they approach the corresponding restoring
frequencies of SLG, $\omega_{\sigma r}$ and $\omega_{\pi r}$, but
they spread out at finite $q$, forming two quasibands.
\cite{Giuliani_1983, Jain_1985} In Fig.~\ref{Fig3} we show such a
quasi-band for $N=5$ in the $\sigma+\pi$ group (the corresponding
quasi-band in the $\pi$ group is not shown to avoid cluttering of
curves in that group). Similar quasibands may be seen in the case of
MWCNTs with, e.g., $N=30$ walls in Fig.\ 4(a) of
Ref.~\cite{Yannouleas_1996} based on a single-fluid model for
$\sigma$ electrons, and $N=10$ walls in Fig.\ 1 of
Ref.~\cite{Chung_2007} based on a two-fluid model for both $\sigma$
and $\pi$ electrons. One sees in Fig.~\ref{Fig3} that the four lower
curves in the $\sigma+\pi$ quasi-band for $N=5$ are quadratically
dispersing, whereas the uppermost curve is linearly dispersing with
a slope that is about five times the slope of the corresponding
$N=1$ curve for small $q$. It can be shown that, by setting
$\cos\left(\omega d/v_{\perp}\right)\approx 1$ as in the case with
$N=2$, the uppermost curve in the $\sigma+\pi$ quasi-band for $N=5$
gives the dominant contribution to the spectrum $P_5(\omega)$. The
large initial slope of that dispersion curve, along with its
tendency to saturate for higher $q$ values, gives rise to a
high-energy tail in $P_5(\omega)$ that forms a bump at about 27 eV,
which is clearly seen in the experimental spectrum for $N=5$ in
Fig.~\ref{Fig1}. As $N$ further increases, this saturation of the
dispersion curves within a quasi-band becomes more prominent because
of a well-defined upper bound, \cite{Yannouleas_1996} which is best
illustrated by resorting to the eigenmode analysis of an infinite
periodic lattice of SLGs.

In the limit $N\rightarrow\infty$ we recover the case of HOPG, which
exhibits two bands of dispersion relations corresponding to the
$\sigma+\pi$ and $\pi$ bulk plasmon modes. They are obtained by
letting $0\le k\le \pi/d$ and solving the equation in Eq.\
(\ref{infinite}) with $\chi_0(q,\omega)$ from the two-fluid model of
Eq.~(\ref{chi}) where $\gamma_\sigma=\gamma_\pi\rightarrow 0$.
\cite{Giuliani_1983, Jain_1985,Shung_1986,Gumbs_1988} The upper and
the lower edges of those bands are defined by $k=0$ and $k=\pi/d$,
respectively, and the corresponding dispersion relations are given
by $\omega=\omega_\pm^\mathrm{up}(q)$ and
$\omega=\omega_\pm^\mathrm{low}(q)$ with ``+'' for the $\sigma+\pi$
plasmon band and ``-'' for the $\pi$ plasmon band, where expressions
in Eq.~(\ref{pm}) are to be used with the frequencies
$\Omega_{\nu}(q)$ in Eq.~(\ref{Omega}) replaced by
$\Omega_{\nu}^\mathrm{up}(q)=\Omega_{\nu}(q)\sqrt{\coth(qd/2)}$ and
$\Omega_{\nu}^\mathrm{low}(q)=\Omega_{\nu}(q)\sqrt{\tanh(qd/2)}$.
Those band edges are shown by the four thick solid (orange) curves
in Fig.~\ref{Fig3}, which all exhibit parabolic dispersions in the
long wavelength limit.

When $q\rightarrow 0$, one can show that the \textit{lower} band
edges in Fig.~\ref{Fig3} approach the restoring frequencies of SLG,
$\omega_+^\mathrm{low}(0)=\omega_{\sigma r}$  and
$\omega_-^\mathrm{low}(0)=\omega_{\pi r}$, whereas the
\textit{upper} band edges approach the values
$\omega_\pm^\mathrm{up}(0)$ that are given by $\omega_\pm$ in
Eq.~(\ref{pm}) when the frequencies $\Omega_{\nu}(q)$ in
Eq.~(\ref{Omega}) are replaced by $\Omega_{\nu p}=\sqrt{4\pi
e^2N_{\nu}^0/m_{\nu}^*}$, corresponding to the bulk plasmon
frequency in the $\nu$th electron fluid with an effective volume
density of $N_{\nu}^0\equiv n_{\nu}^0/d$. Using our parameters from
Fig.~\ref{Fig1}, we find $\Omega_{\sigma p}\approx$ 21.7 eV and
$\Omega_{\pi p}\approx$ 12.5 eV, which are quite close to the
plasmon frequencies listed in the Table 2 of Ref.~\cite{Lucas_1995}
for the in-plane dielectric function of the ADS model. Hence, we
obtain for the long wavelength limits of the upper edges of the bulk
$\sigma+\pi$ and $\pi$ plasmon bands in Fig.~\ref{Fig3} the values
$\omega_+^\mathrm{up}(0)\approx 27.7$ eV and
$\omega_-^\mathrm{up}(0)\approx 7$ eV, respectively, which are
remarkably close to the peak positions seen in the $P_N(\omega)$
curves in Figs.~\ref{Fig1} and \ref{Fig4} for $N>10$. This may be
rationalized by referring to the $N\rightarrow\infty$ limit of the
LEG model, given in Eq.~(\ref{P_bulk}), where we see that the
smallness of the factor $kd=\omega d/v_\perp$ in $S(q,k)$, cf.\
Eq.~(\ref{S}) implies that the dominant spectral weights in
$P_\infty(\omega)$ come from the upper edges (defined by $k=0$) of
the bulk $\sigma+\pi$ and $\pi$ plasmon bands of HOPG.

Hence, for finite $q$, one expects that the peak positions of both
the $\pi$ and $\sigma+\pi$ plasmons in the EEL spectra of HOPG
should closely follow the dispersion relations of the upper edges of
the corresponding plasmon bands. This is corroborated in
Fig.~\ref{Fig3} by a comparison with the peak positions taken from
Figs.~3 and 5 of Ref.~\cite{Kramberger_2010}, which were deduced
from the experimental EEL spectra of HOPG. (We note that the
overdispersion of the $\sigma+\pi$ upper band edge relative to the
experimental points at larger $q$ values in Fig.~\ref{Fig3} may be
reduced if a finite damping rate is introduced in the two-fluid
model.) In addition, we display in Fig.~\ref{Fig3} the peak
positions obtained in Ref.~\cite{Marinopoulos_2004} from \textit{ab
initio} calculations for graphite, showing good agreement with both
the experimental data \cite{Kramberger_2010} and the upper edges of
the plasmon bands. Therefore, the previously mentioned association
of the plasmon peaks in the EEL spectra of thick samples of MLG or
HOPG with the bulk plasmon modes of graphite
\cite{Eberlein_2008,Carbone_2009a} should be made more specific by
stressing that it is the upper edges of the bulk plasmon bands that
give the prevalent contributions to the EEL spectra of such
structures.

On the other hand, for incident electrons at lower speeds, one
expects that increased contributions in the plasmon spectra of HOPG
may also come from the portions of the bulk plasmon bands with lower
energies. In that context, we mention that Fig.~1 in
Ref.~\cite{Baitinger_2006} shows the results of a HREELS experiment
on HOPG using a 200 eV incident electron beam, where contributions
from the entire $\sigma+\pi$ plasmon band are seen to strongly
depend on the electron reflection angle. In particular, a prominent
peak was observed in the spectra at about 13 - 14 eV that did not
move much when the reflection angle was changed.
\cite{Baitinger_2006} While that peak was associated by the author
with a surface plasmon, its presence may be alternatively explained
as resulting from a van-Hove type of singularity in the plasmon
density of states at the lower band edge of the $\sigma+\pi$ plasmon
band. \cite{Morawitz_1993} Clearly, further investigation into the
features of high-energy spectra of HOPG in HREELS is warranted.

\begin{figure}
\centering
\includegraphics[width=0.38\textwidth]{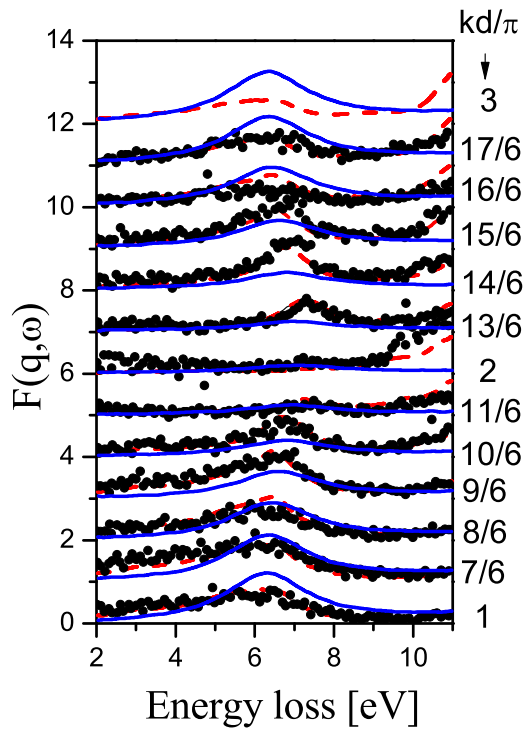}
\caption{(Color online.) Dynamic structure factor [solid (blue) lines] is evaluated from Eq.~(\ref{DSF}) and appropriately scaled to compare it
with the corresponding experimental IXS data (points) from Ref.~\cite{Hambach_2008} with a fixed in-plane wavenumber of $q=$ 0.37 \AA$^{-1}$ for
the indicated values of the perpendicular wavenumber $k$. Also shown are the results of \textit{ab initio} calculations from
Ref.~\cite{Hambach_2008} [dashed (red) lines].}
\label{Fig5}
\end{figure}

It is worthwhile mentioning that the $\pi$ plasmon band was probed
in a recent experiment by Hambach \textit{et al.},
\cite{Hambach_2008} who used IXS to measure the frequency dependent
dynamic structure factor (DSF) of HOPG for momentum transfers beyond
its first Brillouin zone. The experimental data from Fig.~3(a) in
Ref.~\cite{Hambach_2008} are reproduced in our Fig.~\ref{Fig5},
along with the results from their \textit{ab initio} calculations,
showing the frequency dependence of a region near the $\pi$ plasmon
peak for a range of perpendicular wave numbers, $\pi\le kd\le 3\pi$,
with a fixed in-plane wave vector component of about 0.37
\AA$^{-1}$. In particular, the peak position is seen to oscillate
between the values of about 6 and 7 eV, which are commensurate with
the extent of the $\pi$ plasmon band in our Fig.~\ref{Fig3} when one
sets $q=$ 0.37 \AA$^{-1}$ and invokes graphene's nearly isotropic
in-plane polarizability. Furthermore, one can show that the DSF of
an infinite periodic lattice is proportional to the factor
\begin{equation}
F(q,\omega)= \Im\!\left[ \frac{ V(q)\chi_0(q,\omega) }{1+S\!\left(q,k\right) V(q)\chi_0(q,\omega) } \right],
\label{DSF}
\end{equation}
which appears in Eq.~(\ref{P_bulk}) with $S(q,k)$ defined in
Eq.~(\ref{S}). By using the same parameters as those used in
Fig.~\ref{Fig1}, one can easily evaluate from Eq.~(\ref{DSF}) a set
of spectral curves for HOPG with $q=$ 0.37 \AA$^{-1}$ for a range of
$kd$ values, which are seen in Fig.~\ref{Fig5} to reproduce all the
main features of the corresponding experimental data and \textit{ab
initio} results for DSF from Ref.~\cite{Hambach_2008}, including the
$\pi$ plasmon peak position variation with $kd$, and its near
disappearance for $kd=2\pi$. We take that this comparison
demonstrates both the validity of the concept of plasmon bands in
HOPG and the ability of the LEG model to describe them in a simple
analytical manner.

Regarding the applicability of the $N\rightarrow\infty$ limit of the
LEG model in the context of EELS in STEM, we obviously require
$\omega T_N\gg 1$ because, in view of the strong peaking of the
kinematic factor at $q=\omega/v_\perp$ in $P_\infty(\omega)$ from
Eq.~(\ref{P_bulk}), this condition amounts to $Ndq\gg 1$, which
guarantees that the end effects are negligible. When $\omega T_N\sim
1$, the interference is strong and the end effects in such MLG may
not be neglected, making it necessary to use $P_N(\omega)$ from
Eq.~(\ref{P}), even when $N\gg 1$. To illustrate this point, we
mention that, when attempting to model in Fig.~\ref{Fig1} the
experimental EEL spectra for the thickest MLG from
Ref.~\cite{Eberlein_2008} with ``$>\!\!10$'' layers, we could not
obtain nearly as good a fit with $P_\infty(\omega)$ as we did with
$P_N(\omega)$ for $N=13$. The fact that the interference factor in
this case takes the value $\omega T_{13}\approx 1$ points to a need
to look elsewhere for the experimental EEL spectra of thick enough
MLG, where we may test the applicability of $P_\infty(\omega)$ from
Eq.\ (\ref{P_bulk}) to HOPG.

\begin{figure}
\centering
\includegraphics[width=0.48\textwidth]{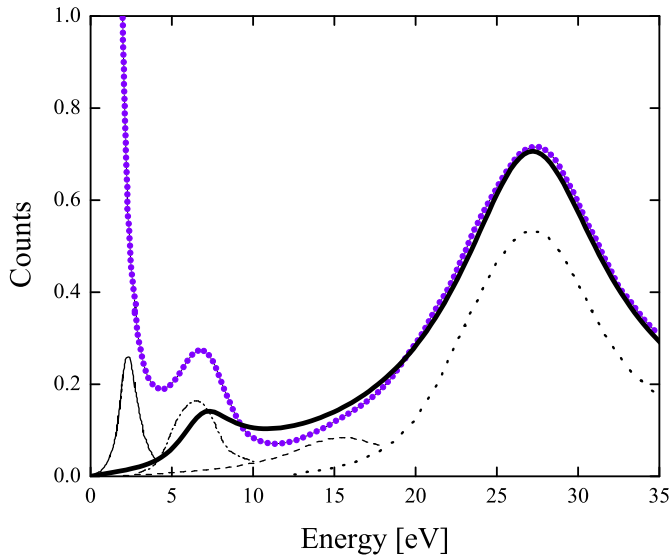}
\caption{(Color online.) The EEL spectra versus energy loss (in eV), evaluated from $P_\infty(\omega)$ in Eq.~(\ref{P_bulk}) [thick (black)
solid line], along with the corresponding experimental spectrum of HOPG from Ref.~\cite{Carbone_2009a} [the (violet) chain curve], which is
theoretically decomposed into a laser-generated peak [thin (black) solid line], $\pi$ plasmon peak [thin (black) dash-dot line],
a ``surface'' $\sigma+\pi$ plasmon peak [thin (black) dashed line], and a ``bulk'' $\sigma+\pi$ plasmon peak [thin (black) dotted
line].}
\label{Fig6}
\end{figure}

In that context, we find it instructive to model the recent
experiment by Carbone \textit{et al.}, \cite{Carbone_2009a} who used
their UEM in the so-called static mode to obtain the EEL spectra of
a slab of HOPG, having such a thickness that $\omega T_N\approx 20$.
In Fig.~\ref{Fig6} we compare the experimental curve from Fig.\ 2 of
Ref.~\cite{Carbone_2009a} for HOPG with our result for the
(appropriately scaled) probability density $P_\infty(\omega)$ from
Eq.~(\ref{P_bulk}), where we used the two-fluid model for
$\chi_0(q,\omega)$ given in Eq.~(\ref{chi}). In order to achieve the
best fit with the experiment, we used the same parameters for the
$P_\infty(\omega)$ curve in Fig.~\ref{Fig6} as those used for the
$P_N(\omega)$ curves in Figs.~\ref{Fig1} and \ref{Fig4}, except for
$\gamma_\sigma$, which was increased to $\gamma_\sigma=9.52$ eV to
take into account that the damping rate for $\sigma$ electrons may
be increased due to more abundant impurities or defects in a thick
slab of HOPG than in a few-layer MLG. Our curve for
$P_\infty(\omega)$ is seen in Fig.~\ref{Fig6} to reproduce the
experiment very well for energies $\gtrsim 10$ eV due to the
increased parameter $\gamma_\sigma$, whereas the $\pi$ plasmon peak
is only qualitatively reproduced in shape, but its position fits the
experiment well. The discrepancy between our model and the
experiment at energies $\lesssim 10$ eV may be partially ascribed to
the presence of both the zero-loss peak and the photogenerated
charge carrier plasma excitations at around 2.4 eV, which were not
subtracted from the experimental curve in Ref.~\cite{Carbone_2009a}.
We also show in Fig.~\ref{Fig4} a phenomenological decomposition of
the experimental curve into four theoretical curves, which was
performed in Ref.~\cite{Carbone_2009a}. Those curves describe
contributions from several processes that were associated by the
authors with: the laser-generated region peaked at about 2.4 eV, the
$\pi$ plasmon peaked at about 7 eV, the surface $\sigma+\pi$ plasmon
peaked at about 15 eV with a low weight, and the bulk $\sigma+\pi$
plasmon peaked at about 27 eV with a much larger weight than its
surface counterpart. \cite{Carbone_2009a} While such a decomposition
of the $\sigma+\pi$ plasmon contributions is corroborated by the
above analysis, we believe that it would be more adequate to label
the peaks at 15 and 27 eV as contributions related to the lower edge
(which carries a signature of the SLG's $\sigma+\pi$ plasmon mode)
and the upper edge of the bulk $\sigma+\pi$ plasmon band in HOPG,
respectively.

\section{Concluding remarks}

We have shown that a simple and straightforward implementation of
the LEG model \cite{Fetter_1974,Giuliani_1983,Jain_1985} to MLG
gives analytical expressions for the spectra of the energy lost by a
fast electron under normal incidence that provide an explicit
account of the effect of increasing number $N$ of graphene layers on
the high-energy plasmon excitations in MLG. Those expressions are
given in terms of a single-layer polarizability $\chi_0(q,\omega)$
that may be modeled independently from the LEG model. By adopting a
phenomenological, two-dimensional, two-fluid model for
$\chi_0(q,\omega)$, which includes Lorentz parameters suitable for
describing the dominant $\pi\rightarrow\pi^*$ and
$\sigma\rightarrow\sigma^*$ inter-band transitions in SLG, we found
good agreement with the plasmon spectra from four independent
experiments. In particular, the shape of such spectra was well
reproduced for MLG with $N\lesssim$ 10, where the most dramatic
change occurs in the $\pi$ and $\sigma+\pi$ plasmon peaks.
\cite{Eberlein_2008} In addition, it was shown that the experimental
EEL spectra for both MLG with $N>10$ \cite{Eberlein_2008} and a
thick slab of HOPG \cite{Carbone_2009a} may be well reproduced by
the same model by allowing for increased damping rates in
$\chi_0(q,\omega)$.

Specifically, the LEG model reproduced the experimentally determined
peak positions, \cite{Eberlein_2008} which were found to move from
about 5 to about 7 eV for the $\pi$ plasmon and from about 15 to
about 27 eV for the $\sigma+\pi$ plasmon as the number of layers in
MLG grows from $N\sim 1$ to $N>10$. \cite{Eberlein_2008} By
referring to the eigenmode analysis of the underlying MLG
structures, both the plasmon peak positions and the experimentally
observed similarity among the EEL spectra for $N=1,2,3,\ldots$ were
found to be derived from the plasmon spectra of SLG, based on the
smallness of the factor $\omega d/v_\perp$ under typical
experimental conditions (low loss EELS and the 100 keV incident
electrons with $d\approx$ 3.35 \AA$^{-1}$). \cite{Eberlein_2008}
Hence no association with the surface plasmons of graphite seems to
be justified for the peaks at about 5 and about 15 eV for the $\pi$
and $\sigma+\pi$ plasmons, respectively, in MLG with few layers. In
addition, the high-frequency tails in the experimental EEL spectra
were exactly reproduced by the LEG model for $N=1$ and 2, and were
identified as arising from the plasmon dispersion and from
integration over a large range of the in-plane wave numbers,
commensurate with the experimental conditions. \cite{Eberlein_2008}

On the other hand, for sufficiently thick MLG, such that
$Nd\omega/v_\perp\gg 1$, the limit of an infinite periodic lattice
of SLGs showed that the peaks at about 7 and about 27 eV come
predominantly from the upper edges of the $\pi$ and $\sigma+\pi$
plasmon bands in the bulk of HOPG, respectively, again because of
the smallness of the factor $\omega d/v_\perp$. This seems to be
corroborated by a comparison with the experimental dispersion
relations found from the EEL spectra of graphite.
\cite{Kramberger_2010} Moreover, it was shown that the
$N\rightarrow\infty$ limit of the LEG model also gives good account
of a dynamic structure factor that was measured by IXS of HOPG,
\cite{Hambach_2008} reproducing the movement of the $\pi$ plasmon
peak between the corresponding $\pi$ plasmon band edges.

When $Nd\omega/v_\perp\sim 1$, the $\sigma+\pi$ plasmon peak is seen
to evolve through a sequence of broad structures between 15 and 27
eV, \cite{Eberlein_2008} which result from the development of a
quasiband of plasmon dispersion curves for MLG with $N\sim 10$ that
is a precursor to a fully developed bulk $\sigma+\pi$ plasmon band
of HOPG. Furthermore, having in mind that the underlying
semi-infinite lattice of SLGs for $N\rightarrow\infty$ does not
support surface states in HOPG placed in the air or vacuum,
\cite{Giuliani_1983, Jain_1985} one may conclude that any remaining
trace of the peak at about 15 eV for the $\sigma+\pi$ plasmon in the
EEL spectra of HOPG is likely to be related to the lower edge of its
bulk $\sigma+\pi$ plasmon band, \cite{Carbone_2009a} and not to a
surface plasmon of graphite.

Given that small variations in the free parameters, used in the
model adopted for $\chi_0(q,\omega)$, do not corrupt the agreement
found with the four independent experiments,
\cite{Eberlein_2008,Carbone_2009a,Hambach_2008,Kramberger_2010} we
suggest that using the LEG model for high-energy plasmon excitations
in MLG, HOPG, and other multilayer carbon nanostructures indeed
presents a versatile and robust alternative to other theoretical
approaches. The novel applications of these analytical models,
presented in this work shed light on several aspects of the problem
at hand, including (a) the role of plasmon dispersion in the spectra
integrated over the wave number, (b) the role of the dynamical
interference factor $Nd\omega/v_\perp$ in determining which
eigenmodes of the underlying MLG structure have prevalent spectral
weights, and (c) the relevance of the bulk plasmon bands, rather
than surface plasmons, in classifying the observed plasmon peak
frequencies.

\begin{acknowledgments}
This work was supported by the Ministry of Education and Science,
Republic of Serbia (Project No.\ 45005). Z.L.M.\ also acknowledges
support from the Natural Sciences and Engineering Research Council
of Canada.
\end{acknowledgments}

\end{document}